\documentclass[twocolumn,showpacs,preprintnumbers,amsmath,amssymb]{revtex4}
\usepackage{graphicx}
\usepackage{dcolumn}
\usepackage{bm}
\usepackage{color}
\begin{document}
\preprint{APS/123-QED}
\title{
Evidence for
gradual evolution of low-energy
fluctuations underlying the first-order structural and valence order in YbPd
}
\author{R. Nakanishi$^1$}
\author{T. Fujii$^1$}
\author{Y. Nakai$^1$}
\author{K. Ueda$^1$}
\author{M. Hirata$^2$}
\author{K. Oyama$^3$}
\author{A. Mitsuda$^3$}
\author{H. Wada$^3$}
\author{T.~Mito$^1$}
 \email{mito@sci.u-hyogo.ac.jp}
\affiliation{
$^1$Graduate School of Material Science, University of Hyogo, Ako-gun 678-1297, Japan\\
$^2$Institute for Materials Research, Tohoku University, Sendai 980-8577, Japan\\
$^3$Graduate School of Science, Kyushu University, Fukuoka 819-0395, Japan
}
\date{\today}

\begin{abstract}
The valence orders at $T_a=125$~K and $T_b=105$~K in
the
cubic compound YbPd have been investigated by $^{105}$Pd-nuclear magnetic resonance (NMR) measurements.
Significant decrease in the density of states at
the
Fermi energy below $T_a$ is evident from the measurement of nuclear spin lattice relaxation rate $1/T_1$, suggesting that the instabilities of
Fermi surface are associated with the transitions.
Moreover we observed the unusual evolution of low-energy fluctuations toward the valence transition at $T_a$ behind its drastic first-order nature.
The structural transition
accompanying
the valence order
may
occur as a result of cooperative effect of Fermi surface and valence instabilities.
\end{abstract}

\maketitle


A rich variety of intriguing properties that lanthanide and actinide compounds show, especially those related to the $f$-electrons degrees of freedom, have provided good opportunities to investigate issues of strong electron-electron correlations.
Among them, the mechanisms of many phenomena involving the valence degrees of freedom have remained unsolved, because it is difficult to observe directly
the dynamics of valence properties, including valence fluctuations.
For example,
the valence fluctuation in the prototypical intermediate valence material SmB$_6$ was {\it indirectly} corroborated by
contrasting appearances
in two experimental results:
Sm$^{2+}$ and Sm$^{3+}$ states are distinguishable in
x-ray absorption spectroscopy \cite{Vainshtein}
but not
in $^{149}$Sm M${\rm \ddot{o}}$ssbauer spectroscopy \cite{Cohen},
implying that the Sm valence fluctuates at time scales between those characteristic of the two experiments.

In this letter, we report $^{105}$Pd-nuclear magnetic resonance (NMR) measurements of valence ordered system YbPd.
Of particular interest is examining the possibility of observing slow valence fluctuations somehow
{\it directly} by the NMR measurement just above the valence ordering temperature.
Generally,
the valence fluctuates much faster than the NMR frequency
[several tens of megahertz].
YbPd,
which crystallizes in the cubic CsCl-type structure,
undergoes unusual phase transitions at $T_a = 125$ K and $T_b = 105$ K,
followed by complicated magnetically ordered states below $T_{\rm N}=1.9$~K \cite{Pott,Bonville,Oyama}.
At the transitions at $T_a$ and $T_b$, which are focused on in the present study,
the specific heat shows sharp peaks at these temperatures, suggestive of first-order nature.
However no anomaly is seen there in the magnetic susceptibility \cite{Iandelli}, and the mean Yb valence is little temperature dependent [2.83 at room temperature] \cite{Pott}.
Interestingly, YbPd
remains
metallic through the transitions, which is unique because known valence ordering systems, such as Yb$_4$As$_3$ and EuPtP, exhibit an increase in the resistivity when the valence ordering takes place.

Recently,
it turned out that
the transitions at $T_a$ and $T_b$ were
accompanied by structural changes,
namely the cubic
structure lowers into
a tetragonal symmetry \cite{Mitsuda,Takahashi}.
Moreover doubling of the unit cell was
found out
below $T_b$
\cite{Mitsuda,Takahashi}, implying the occurrence of valence ordering.
The
ordered
structures proposed to date are as follows;
for $T_{\rm N} <T< T_b$ (LT phase),
Yb$^{3+}$ and Yb$^{2.6+}$ ions
alternately stack
along the $c$-axis,
and for $T_b <T< T_a$ (IT phase),
the structure is of incommensurate characterized by
a wave vector $\left( \pm 0.07 \ \pm 0.07 \ 1/2 \right)$ 
\cite{Takahashi}.
Thus, one expects that the valence degrees of freedom are related to driving force
for
the transitions at $T_a$ and $T_b$,
however there had been no
conclusive
experiment to verify the hypothesis so far.
Here we present unambiguous evidence for critical behavior towards the transition at $T_a$, which clearly reveals the evolution of low-energy correlations underlying the first-order structural transition.

A difficulty in accomplishing the present study is that the intensity of $^{105}$Pd-NMR signal is generally weak due to the small gyromagnetic ratio $\gamma$  ($\gamma/2 \pi = 1.95$ MHz/T) and low natural abundance $\sim 22.2 \%$ for $^{105}$Pd.
Moreover, below $T_b$, effectively nonzero nuclear quadrupole interaction in the tetragonal structure
causes
broadening of the NMR signal, resulting in additional weakening.
Therefore we performed the measurements with the help of high magnetic fields
using a 400~MHz NMR superconducting magnet (SM) (about 9.4~T) and a 15~T SM ($12.1<H<13.7$~T) both in University of Hyogo, and a 25~T cryogen-free SM (25T-CSM) ($21.5<H<24.0$~T) in Institute for Materials Research, Tohoku University.
All the $^{105}$Pd-NMR experiments were carried out by using the spin-echo technique with a phase-coherent pulsed spectrometer.
A high-quality strain-free YbPd
powder sample
was synthesized by a newly developed method
where
a powdered Yb-rich material Yb$_5$Pd$_2$, obtained by a flux method, is annealed to evaporate Yb to YbPd.
The detaileds are described in Ref.~\cite{Oyama}.

\begin{figure}[t]
\includegraphics[width=1\linewidth]{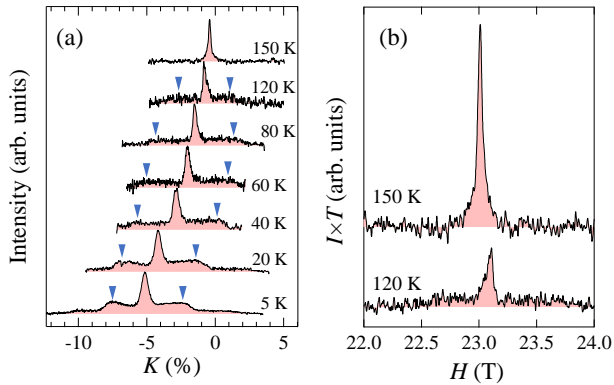}
\caption{\label{fig:epsart}
(color on line)
(a) $^{105}$Pd NMR spectra at different temperatures.
The data
obtained by sweeping magnetic field using 25T-CSM
are shown as a function of the Knight shift, $K$.
Spectral intensity is normalized by
their peak intensity.
The arrows indicate the first satellite of nuclear quadrupole splitting.
The intensity just above and below $T_a$ is compared in (b).
The data were measured at a frequency of 44.68~MHz.
%
}
\end{figure}

Figure 1(a) shows the temperature dependence of $^{105}$Pd NMR spectrum.
The Lorentzian-like line shape at 150~K
in the high-temperature (HT) phase
drastically changes into the so-called powder pattern below $T_a$.
As shown in Fig.~1(b), the intensity of the single line at 150~K is significantly reduced at 120~K, while additional weak signal continuously emerges in the region from 22.5 to 23.8~T.
These results indicate that the high-symmetry
in the HT phase
changes to a lower one below $T_a$:
nuclear quadrupole interaction at a Pd nuclear position, which does not effectively work in the
HT
phase
with the cubic structure,
brings about splitting and broadening of the spectrum.
The symmetry lowering below $T_a$ is well consistent with the occurrence of structural change reported previously \cite{Mitsuda,Takahashi}.
Note that the powder pattern is clearly observed even at very high field above 20~T, indicating that magnetic anisotropy in the LT phase with the tetragonal structure is weak.

From the shoulder structure indicated by the arrows in Fig.~1(a),
identified as the first satellite of
nuclear quadrupole splitting, nuclear quadrupole resonance (NQR) frequency $\nu_{\rm Q}$ is estimated to be 2.3~MHz at 5~K and increases with increasing temperature [$\nu_{\rm Q} = 2.75$~MHz at 80~K].
$\nu_{\rm Q}$ is the physical value that sensitively reflects local charge distribution through the 
electric field gradient (EFG),
and it may be useful to compare the present temperature dependence with the result reported for YbPd$_2$Si$_2$ \cite{Emi} in which the Yb valence uniformly fluctuates.
For YbPd$_2$Si$_2$, the resemblance between the temperature dependences of $\nu_{\rm Q}$ measured at the Pd site and mean Yb valence [both monotonically increase with increasing temperature] simply reveals intimate relationship between the two
physical values.
In the case of the LT phase of YbPd, the situation is more complex, because the Yb ions are in the antiferro-type
valence
ordered state
and the EFG at the Pd site surrounded by eight Yb ions (four Yb$^{3+}$ and four Yb$^{2.6+}$) should be influenced from the both valence states.
The present experimental fact of temperature dependent $\nu_{\rm Q}$ in the LT phase
may suggest that local electronic configurations at each Yb valence still change at low temperatures,
which may be undetectable
by the simple measurement of
mean Yb valence \cite{Pott}.

\begin{figure}[t]
\includegraphics[width=1\linewidth]{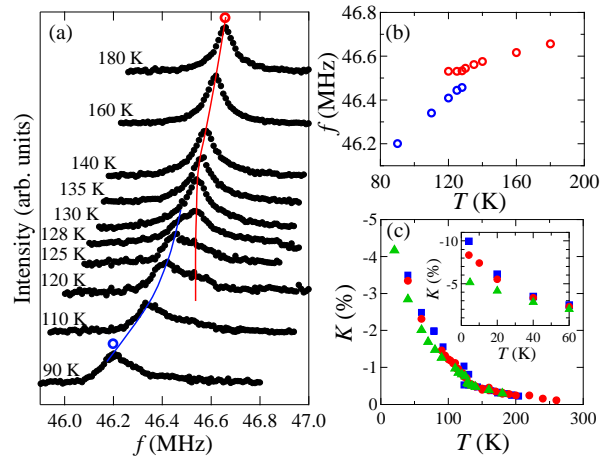}
\caption{\label{fig:epsart}
(color on line)
(a) The detailed temperature dependence of NMR line, which was measured by sweeping the frequency at a constant field of 24.0~T.
The intensity is normalized by the peak value of spectra.
For $T<130$~K, we measured near the central line.
The resonance frequencies for the cubic phase (red circle) and the tetragonal (blue circle) are traced by the red and blue lines, respectively.
(b) Temperature dependences of the resonance frequencies determined in Fig. 2(b).
(c) Temperature dependence of
$K$.
The squares, circles, triangles denote the data measured using 400MHz-SM, 15T-SM, and 25T-CSM, respectively.
The inset shows an expanded view of $K$ below 60~K.
}
\end{figure}

Further detailed spectral features are revealed in Fig.~2.
As temperature decreases below 128~K, another resonance line (blue line) evolves at lower frequencies.
Note that this additional signal
is a central line corresponding to a $-1/2 \leftrightarrow 1/2$ transition.
The two signals arising from the HT and IT phases coexist for $120 \leq T \leq 128$~K from our data.
As demonstrated in Fig.~2(b) where the resonance frequencies are plotted as a function of temperature, the discontinuous shift and the coexistence of the two lines
around
$T_a$ are characteristics of first-order transition.
On the other hand,
at $T_b$, no significant anomalies are seen in the central line as well as in the Knight shift, $K$, as described later.
Instead, the satellite shoulder structure becomes unclear in the LT phase [see the spectrum at 120~K in Figs.~1(a) and(b)], indicating that $\nu_{\rm Q}$ is not uniformly determined, and consistent with the incommensurate structure proposed by Ref. \cite{Takahashi}.
The transition at $T_b$ is observable through the electric interaction rather than the magnetic ones in the NMR measurements.

$K$ estimated from the main peak indicated in Fig.~2(a) is plotted in Fig.~2(c).
The temperature dependence is roughly consistent with that of the susceptibility $\chi$ \cite{Iandelli}:
the absolute value of $K$ increases with decreasing temperature.
However, $K$ exhibits a small but clear jump at $T_a$, which is different from $\chi (T)$ that increases with cooling smoothly.
$K$ is generally decomposed into temperature dependent and independent terms;
\begin{eqnarray}
K = K(T) + K_c.
\end{eqnarray}
Here $K(T)$ is a Curie-Weiss term arising from $4f$ electrons possessing localized characters, and the temperature independent term $K_c$ mainly arises from Pauli paramagnetism, which is enhanced through the Fermi contact interactions of conduction electrons.

\begin{figure}[t]
\includegraphics[width=1\linewidth]{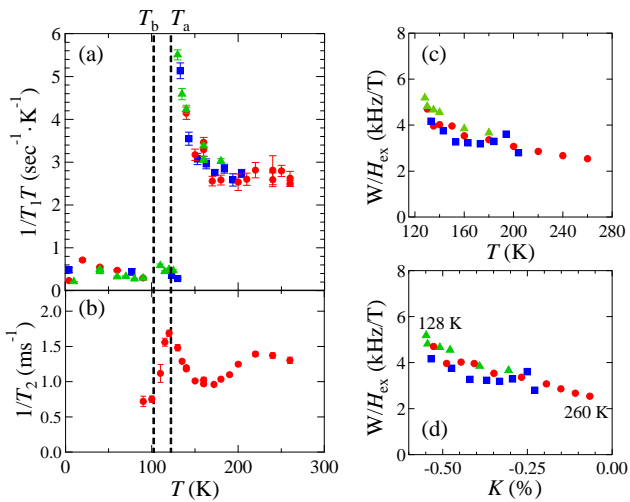}
\caption{\label{fig:epsart}
(color on line)
Temperature dependences of (a) $1/T_1T$, (b) $1/T_2$, and (c) spectral width $W$ divided by
external
field $H_{\rm ex}$.
(d) $W/ H_{\rm ex}$ vs $K$ plot with temperature as an implicit parameter.
}
\end{figure}

We have also measured nuclear spin lattice relaxation time $T_1$ at the resonance peaks shown in Fig.~2(a).
The process of estimating $T_1$ from relaxation curves is described in Ref.~\cite{Suppl}.
Evaluated $1/T_1T$ was found to show unique temperature dependence
as demonstrated in Fig.~3(a):
(i) $1/T_1T$ is almost temperature independent at high temperatures above 160~K
and at low temperatures below $T_a$, and (ii) $1/T_1T$ rapidly increases below 160~K down to $T_a$, followed by a significant drop just below $T_a$.
$1/T_1T$ is generally expressed as follows;
\begin{eqnarray}
\frac{1}{T_1T} = \frac{2\gamma_{\rm n}^2 k_{\rm B}}{(\gamma_{\rm e} \hbar)^2} \sum_{q} \left[ A_{\rm hf}(q)^2
 \frac{{\rm Im} \chi (q, \omega_0)}{\omega_0} \right],
\end{eqnarray}
where $\gamma_{\rm n}$ and $\gamma_{\rm e}$ are the nuclear and electronic gyromagnetic ratios, respectively, $\omega_0$ is the NMR frequency,  $A_{\rm hf}(q)$ is hyperfine coupling constant, and ${\rm Im} \chi (q, \omega_0)$ is the imaginary part of $\chi (q, \omega_0)$.
In weakly correlated electron systems, such as a Fermi liquid state, it is known that $1/T_1T$ becomes little temperature dependent as $1/T_1T \propto \left \{ A_{\rm hf} D(\varepsilon _{\rm F}) \right \}^2$, where $D(\varepsilon _{\rm F})$ is the density of states at Fermi energy $\varepsilon _{\rm F}$.
In the similar manner to
$K$
[see Eq.~(1)],
$1/T_1T$ in the present system is decomposed into
$\left( 1/T_1T \right)_f (T)$ and $\left( 1/T_1T \right)_c$,
where the former is a temperature dependent term arising from the localized $4f$ electrons and the latter is the temperature independent contribution of conduction electrons.
Then, the phenomenon (i) is accounted for in terms of $\left( 1/T_1T \right)_f \ll  \left( 1/T_1T \right)_c$.
As $1/T_1T$ originating from well localized $f$ electrons tends to obey $1/T$, $\left( 1/T_1T \right)_f$ can
be
small at high temperatures, resulting in the temperature dependence dominated by $\left( 1/T_1T \right)_c$.
Indeed, such a situation is realized for example in YbCo$_2$Zn$_{20}$ which is known as an extremely heavy Fermion \cite{Mito}.


In the IT and LT phases, $1/T_1T$ again shows weak temperature dependence,
indicating
that the contribution of
Yb magnetic sites
to the $T_1$ relaxation is weak.
The decrease in $1/T_1T$
when temperature is lowered
from the HT to the LT phases indicates a decrease in $\left \{ A_{\rm hf} D(\varepsilon _{\rm F}) \right \}^2$.
By assuming that $A_{\rm hf}$ is unchanged, the ratio of $1/T_1T$ in the HT and LT phases, $(1/T_1T){\rm (HT)}/ (1/T_1T){\rm (LT)}=9$, gives the rough estimation of $70 \%$ decrease in $D(\varepsilon _{\rm F})$.

For more accurate estimation, one needs to know $A_{\rm hf}$ in the HT and LT phases.
If we consider two main contributions to $A_{\rm hf}$, namely hyperfine fields originating from localized $f$ electrons and conduction electrons, $A_{{\rm hf},f}$ and $A_{{\rm hf},c}$, respectively,
it is easy to estimate $A_{{\rm hf},f}$ from the temperature dependent parts of $K$ and $\chi$.
From the $K$ vs $\chi$ plot, $A_{{\rm hf},f}$ is estimated as $-7.7$~kOe/$\mu_{\rm B}$ in the HT phase
while it is $-5.2$~kOe/$\mu_{\rm B}$ in the LT phase on the basis of the valence order model
that magnetic and nonmagnetic Yb ions alternately stack along the $c$ axis \cite{Takahashi}.
Here, as shown in the inset of Fig.~2(c), one sees strong field influence on $K$ at high fields above 21.5~T and low temperatures below 40~K, which is ascribed to saturating tendency in the magnetization \cite{Sugishima}.
Therefore we estimated $A_{\rm hf}$ using the data measured at the lowest field.
[see Fig.~5 in Ref. \cite{Suppl}].
This reduction in the LT phase is intuitively understood, because $4f$ electrons from the Yb$^{3+}$ ion are more localized than
in the HT phase, so that
transferred hyperfine field at a Pd nucleus position from the $4f$ electrons
will become smaller.

On the other hand, we cannot evaluate $A_{{\rm hf},c}$ giving rise to temperature independent components in question of $K$ and $1/T_1T$.
However, since $A_{{\rm hf},c}$ is dominated by atomic hyperfine coupling with on-site conduction electrons, 
it is unlikely that $A_{{\rm hf},c}$ changes
at $T_a$
as much as $A_{{\rm hf},f}$
does.
(Even though we assume that,
in the LT phase,
$A_{{\rm hf},c}$ is reduced by the same ratio with that of $A_{{\rm hf},f}$, it accounts for only a $50\%$ of the reduction of $1/T_1T$.)
Thus, the significant drop of $1/T_1T$ from the HT and LT phases is attributed to the decrease in $D(\varepsilon _{\rm F})$,
suggestive of Fermi surface instability in the vicinity of $T_a$.
For the IT phase, we cannot proceed to analyze the $1/T_1T$ data, because the complex valence ordering state is proposed \cite{Takahashi} and also we could not obtain accurate data due to a poor signal intensity.
However the large drop of $1/T_1T$, quite analogous to the data in the LT phase, suggests the resultant reduction of $D(\varepsilon _{\rm F})$ as well as in the LT phase.

For another remarkable feature (ii) described above, the rate of the $1/T_1T$ increase with cooling is much faster than $1/T$, indicating that this is critical behavior due to the evolution of any low-energy fluctuations toward the phase transition at $T_a$.
The
similar phenomenon so called `critical slowing down' is often observed when a second-order magnetic 
phase transition
occurs.
However,
in YbPd, almost no magnetic anomaly is found at $T_a$ and the phase transition has been identified as a structural transition accompanied by valence redistribution \cite{Mitsuda,Takahashi}.
Moreover, since magnetic ordering temperature $T_{\rm N} = 1.9$~K is far below $T_a$, 
it is very unlikely that
magnetic correlations associated with the magnetic order develop so steeply at such high temperatures $T>T_a$.
Thus it is convincing that the critical behavior 
is attributed to the phase transition at $T_a$ itself, probably associated with slowing valence fluctuations down.

We discuss how the critical behavior appears in YbPd.
In Fig.~3(b), the temperature dependence of nuclear spin-spin relaxation rate $1/T_2$ is shown.
Although the origin of its anomalous temperature dependence above 180~K is not clear to date, $1/T_2$ shows the divergent behavior
and
the peak is
at slightly lower temperature than that of
$1/T_1T$.
Similar pronounced peaks in $1/T_1$ and $1/T_2$ are reported in the NQR study of a charge order system,  Pr based cuprate PrBa$_2$Cu$_4$O$_8$ \cite{Fujiyama}.
%
%
%
The relaxation anomalies are accompanied with broadening of NQR spectra,
and they arise from
the slow fluctuations of
EFG
due to charge ordering instability,
which freeze randomly at lower temperatures.
In the case of YbPd, the formation of incommensurate Yb valence order is proposed \cite{Takahashi}, so that similar line broadening originating from the slow EFG fluctuations is expected.
Actually,
NMR line gradually broadens as temperature decreases down to $T_a$ as shown in Fig.~3(c).
However,
Fig.~3(d), where the spectral width divided by external field $W/H_{\rm res}$ and $K$ are plotted with temperature as an implicit variable, demonstrates a linear relation between the two values.
This implies
that
a magnetic origin rather than an electric one is responsible for
the broadening.
Once the perturbation of nuclear quadrupole interaction works on nuclear spin states through the effective EFG, the NMR line immediately split off to satellite peaks such as indicated by the arrows in Fig.~1(a).
Therefore the present $T_1$ measurement at the remaining main peak is not sensitive to electronic interactions.
Note that
this
conclusion that the critical behavior in $1/T_1T$ appears through magnetic mechanisms does not mean the absence of charge fluctuations.

The valence fluctuations of Yb ion between Yb$^3+$ and Yb$^2+$
should
simultaneously bring about magnetic fluctuations due to the difference of magnetic characters between the two ionic states: the Yb$^3+$ state is magnetic and Yb$^2+$ state is nonmagnetic.
Therefore
one of
the plausible interpretations of the 
critical behavior in the relaxation data
is that our measurement selectively detected the magnetic fluctuations by observing the NMR signals insensitive to electric interactions.
Such magnetic fluctuations induced by the valence fluctuations 
has not been
observed so far to our knowledge.
We suppose that the very unique valence order of YbPd is one of crucial points for the
successful
observation.
Namely, compared to valence fluctuations between $2+$ and $3+$ states, the small difference between the two valence states, $3+$ and noninteger valence ($2.6+$), should be accompanied with low-energy magneatic fluctuations that one can easily detect by NMR measurements.
[For example, in the case of EuPtP which undergoes two valence orders with integer valences (2+ and 3+) at 240 and 200~K, no critical behavior is observed in the $T_1$ measurement at all \cite{Mito}.]

In this study, the first order nature of the phase transition at $T_a$ is evident from the measurement of NMR line [see Fig.~2(b)].
Besides the reduction of $D(\varepsilon _{\rm F})$ caused by the phase transitions, indicated by the $T_1$ measurement [see Fig.~3(a)], suggests
that the phase transition
is
driven by a Fermi surface instability.
These observed features
are consistent
with the idea of
conventional
band Jahn-Teller effect \cite{Mitsuda},
reminiscent of a cubic-to-tetragonal structural transition in CsCl-type structure compounds LaAg$_{1-x}$In$_x$ and $R$Cd ($R=$ La, Ce, and Pr) \cite{Balster,Hasegawa,Kurisu,Aleonard}.
Indeed, a band structure calculation based on the density functional theory within fully relativistic scheme suggests that the spin-orbit interaction splits the $4f$ states into two manifolds and one of them ($4f_{7/2}$) forms
the large density of states
near the Fermi level \cite{Jeong}.

However,
this scenario does not
cover all
the remarkable properties of the phase transition at $T_a$, in particular the
occurrence
of valence ordering.
In this context,
the most interesting
finding
in the present study
will provide important clue to comprehensive clarification:
namely, behind the drastic first order structural transition, low-energy interactions associated with the valence order gradually  evolve.
The structural transition
involving
valence redistribution in YbPd may be captured as the cooperative effect of the instabilities of the Fermi surface and the valence.


In summary, the use of static high magnetic fields up to 24.1 T enabled the precise $^{105}$Pd-NMR measurement on YbPd.
The discontinuous line shift at $T_a$ and the sudden appearance of nuclear quadrupole split line below $T_a$ are consistent with the cubic-to-tetragonal structural transition with first-order nature.
Below $T_a$, the significant reduction of $D(\varepsilon _{\rm F})$ is also evident from the  measurement of $1/T_1T$.
These results suggest
that Fermi surface properties, including band Jahn-Teller effect, play an important role in the phase transition.
Moreover we observed the gradual evolution of low-energy interactions toward the valence order at $T_a$
behind the first order structural transition.
This observation is interpreted as the detection of magnetic fluctuations induced by the valence fluctuations simultaneously.
The structural transition accompanied with the valence order seems to occur as a result of cooperative effect of these two factors.

\begin{acknowledgments}
This work was supported by JSPS KAKENHI (Grants No. 16K05457, No. 18H04331, and No. 15H05883).
The experiments at high magnetic fields above 20~T were performed under the Inter-University Cooperative Research Program of the Institute for Materials Research, Tohoku University (Proposal No. 18H0024).
\end{acknowledgments}



\end{document}